\documentclass[twocolumn,showpacs,preprintnumbers,amsmath,amssymb,revsymb,prl, superscriptaddress]{revtex4}
\usepackage[dvips]{graphicx}
\usepackage{mathptmx}
\usepackage{bm}

\newcommand{\lbrla}{\left\langle }
\newcommand{\rbrra}{\right\rangle }

\newcommand{\lbrp}{\left| }
\newcommand{\rbrp}{\right| }
\newcommand{\D}[2]{\frac{d #1}{d #2}}

\newcommand{\bra}[1]{\lbrla #1 \rbrp}
\newcommand{\ket}[1]{\lbrp #1 \rbrra}
\newcommand{\braket}[2]{\lbrla #1 | #2 \rbrra}

\begin{document}
\title{Stroboscopic wavepacket description of non-equilibrium many-electron problems}
\author{P. Bokes } \email{peter.bokes@stuba.sk}
\affiliation{Department of Physics, University of York, Heslington, York
         YO10 5DD, United Kingdom}
\affiliation{Department of Physics, Faculty of Electrical Engineering and
        Information Technology, Slovak University of Technology,
    Ilkovi\v{c}ova 3, 812 19 Bratislava, Slovak Republic}
\author{F. Corsetti}
\author{R. W. Godby}
\affiliation{Department of Physics, University of York, Heslington, York
         YO10 5DD, United Kingdom}

\date{\today{}}

\begin{abstract}
We introduce the construction of a orthogonal wavepacket basis set, using the concept 
of stroboscopic time propagation, tailored to the efficient description 
of non-equilibrium extended electronic systems. Thanks to three desirable properties 
of this basis, significant insight is provided into non-equilibrium processes (both 
time-dependent and steady-state), and reliable physical estimates of various many-electron
quantities such as density, current and spin polarization can be obtained. The use 
of this novel tool is demonstrated for time-dependent switching-on of the bias in quantum
transport, and new results are obtained for current-induced spin accumulation at 
the edge of a 2D doped semiconductor caused by edge-induced spin-orbit interaction.
\end{abstract}


\maketitle

Wavepackets (WP) are a very useful concept when analyzing quantum mechanical 
scattering processes, since they combine local and wave-like aspects on an equal footing. 
Some of their more recent applications range from studies of the intrinsic 
spin Hall effect in semiconductors\cite{Culcer04,Nikolic05}, spin-flip 
dynamics\cite{Kim05}, thermal averaging and its influence on interference 
patterns\cite{Heller05} or transport of an electron through Luttinger 
liquid\cite{LeHur06}. However, the use of traditional WPs in degenerate fermionic 
systems raises difficulties since the exclusion principle restricts the available 
eigenstates that are superposed within a single WP. 
%
Several orthogonal wavepacket\cite{Stevens83,Yamada05,Chen94} and wavelet\cite{Wei95} 
approches were put forward in the past to accomodate the exclusion principle;
however in contrast to our WPs these do not directly relate to typical many-electron 
states such as the electronic ground state or moderate perturbations from it 
at zero temperature.

If we forego the time-dependent feature of WPs, the latter problem is conveniently 
resolved with the introduction of Wannier functions\cite{Wannier37,Marzari97}: 
by occupying a finite number of them, we locally recover the exact eigenstates 
of a system of non-interacting electrons. 

In this work we combine the advantages of Wannier functions for extended systems 
with the time-dependent description of WP propagation. This is achieved by generalizing 
the orthogonal WPs introduced by Martin and Landauer\cite{Martin92} for ideal 1D leads.  
Our wave-packet basis set (WPB) has following three properties: 
(1) each basis function (WP) is localized in space, 
(2) occupying a subset of the WPB we recover the exact non-interacting many-electron 
ground state of a reference Hamiltonian, 
(3) the WPB is generated by time propagation through successive time-steps, $\tau$, 
of {\it an initial set} of WPs, according to a reference Hamiltonian.

\begin{figure}[b]
\includegraphics[width=8cm]{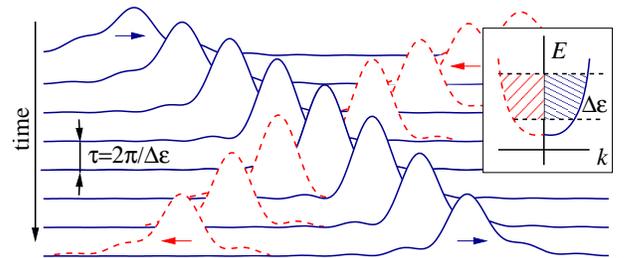}
\caption{Squared amplitudes of orthogonal stroboscopic wavepackets obtained by time 
propagation of the initial WP (in the centre) by a constant time-step $\tau$. 
The right- (blue/full) and left- (red/dashed) WPs belonging from the same energy 
band (inset) are shown. These, together with the WPs coming from the bands covering 
the rest of the spectrum, form a complete orthogonal basis set.} \label{fig-1}
\end{figure}

From the above properties it follows that we can view
the whole basis set as a {\it stroboscopic pictures} of a continuous time-evolution 
of a suitably chosen family of {\it initial} WPs (Fig.~\ref{fig-1}). 
Since all WPs are orthonormal, each copy can be occupied by precisely one electron 
and in time $\tau$ each electron will move into 
its neighboring' WP. Similarly, if a single electron is in a superposition of several
WPs, in time $\tau$ it will be in the {\it same} superposition but of the WPs obtained
from the former by a single shift of the basis functions. This picture 
is valid as long as the reference Hamiltonian is time-independent in the region 
where the concerned WPs are localized. We will refer to this region as the {\it bulk}
and to the rest, typically a much smaller region than the bulk, as the {\it scatterer}.
Similarly, the bulk (scattering) WPs are those WPs that are generated with the bulk
(bulk+scatterer) Hamiltonian. 

To obtain the time-dependent dynamics in the scatterer one needs to perform 
a full time-dependent simulation of the bulk WPs entering the scatterer. 
After certain time, 
the scattering WPs will return into the bulk 
where 
those WPs can once again be expanded into the bulk WPB 
and propagated as moves of duration $\tau$ between the bulk WPs, i.e. analytically.
Hence, the WPB offers a very simple interpretation of the processes 
as well as a framework to perform numerical time-dependent simulations.

The consistency of the conditions (1) and (3) demands that the reference 
Hamiltonian posses translational symmetry in the direction of propagation.
Its eigenstates in the Bloch form will be sufficient to create a basis such that each 
WP from the initial set will be spatially localized and their time-propagated 
WPs will slowly disperse with increasing time. This property can be satisfied only 
if the reference Hamiltonian is just that of the bulk. We may also construct 
the WPB for the combined system where the reference Hamiltonian is that 
of bulk+scatterer, but 
the scattering will typically result in a strongly delocalized WP (e.g.
transmitted and reflected components).
However, the scattering-WPs can be easily expanded into the bulk WPB, 
a fact of which we will make use later.

{\bf Definition of the WPB and its formal properties}.
To define the basis set let us take an extended system specified by 
the reference Hamiltonian $\hat{H}$ with a continuous spectrum of 
eigenenergies $\epsilon \in (\epsilon_0,\infty)$,
$
	\hat{H} \ket{\epsilon,\alpha} = \epsilon \ket{\epsilon,\alpha}.
$
To each eigen energy we will generally have a set of degenerate single-particle 
eigenstates 
$\ket{\epsilon,\alpha}, \quad \alpha=1,2,...,N_{\epsilon}$, 
forming all together a complete orthogonal whose normalization we choose such that 
$
	\braket{\epsilon',\alpha'}{\epsilon,\alpha} =  \label{eq-f1}
			\delta(\epsilon-\epsilon') \delta_{\alpha,\alpha'}.
$

From the above set we can generate an orthogonal and complete wave-packet 
basis set (WPB) by first choosing the {\it initial set} of wave-packets\footnote{We 
assume that the number of degenerate states, $N_\epsilon$, is the same for all 
energies $\epsilon' \in (\epsilon^\alpha_n,\epsilon^\alpha_{n+1})$.}
\begin{equation}
	\ket{n,0,\alpha} =  \label{eq-f2}
	\frac{1}{\sqrt{\Delta \epsilon_n}} 
	\int_{\epsilon^\alpha_n}^{\epsilon^\alpha_{n+1}} d \epsilon'
	U_{\alpha,\alpha'}(\epsilon')\ket{\epsilon',\alpha'}, \quad n=0,1,2,...
\end{equation}
for an arbitrarily chosen division of the spectrum into {\it energy bands} 
$\left\{ (\epsilon^\alpha_n,\epsilon^\alpha_{n+1}) \right\}_{n=0}^{\infty},
\quad \alpha=1,2,\ldots,N_\epsilon$ with bandwidths 
$\Delta \epsilon^\alpha_n=\epsilon^\alpha_{n+1}-\epsilon^\alpha_{n}$. 
The division into energy bands 
must cover the full spectrum of $\hat{H}$ but otherwise can be chosen 
so as to suit the physical situation as discussed later.
The unitary, energy-dependent matrix $U_{\alpha \alpha'}(\epsilon)$ 
represents the second freedom of choice in the construction of the WPB.
In this work we will use 
$U_{\alpha \alpha'}(\epsilon) = \delta_{\alpha \alpha'}$ which is satisfactory 
for our present purposes, but in general it can be used either to adopt 
the bulk WPB to the scattering processes involed i.e. by shifting WPs 
in cetrain bands ($U_{\alpha \alpha'}(\epsilon) = \delta_{\alpha \alpha'} e^{-ik_x x_0}$),
or to improve the localization of the WPs, in analogy with Wannier 
functions\cite{Marzari97}.
All the functions $\left\{ \ket{n,0,\alpha} \right\}_{n}$ are orthogonal by definition, 
since they are linear combinations of eigenstates from disjunct energy bands.

The construction of the WPB is completed by forward and backward time 
propagation of the initial set 
\begin{equation} 
	\ket{n, m,\alpha} =  \label{eq-f3}
	e^{-i\hat{H} m \tau_n} \ket{n,0,\alpha}, \quad
	m = \pm 1, \pm 2, \ldots 
\end{equation}
by regular, band-dependent time steps 
$\tau^\alpha_n=2\pi / \Delta \epsilon^\alpha_n$.
It is easy to verify that this choice of time step {\it guarantees orthonormality 
of consecutive wave-packets within each band}
\begin{equation} 
	\braket{n,m,\alpha}{n,m',\alpha} =  \delta_{m,m'}.
\end{equation}

Due to the orthogonality of the WPs we can uniquely expand any eigenstate 
of the reference Hamiltonian into the WPB with expansion
coefficients 
$ \braket{\epsilon,\alpha}{n, m,\alpha}
	= (\Delta \epsilon_n)^{-1/2} \exp\{-i\epsilon m \tau_n\}$, 
with $\epsilon \in (\epsilon_n,\epsilon_n+\Delta \epsilon_n)$.
Conversely, combining this with Eqs.\ref{eq-f2} and \ref{eq-f3} one obtains that 
$
	\sum_m \ket{n, m, \alpha}
	\braket{n, m, \alpha}{\epsilon,\alpha}
	= \ket{\epsilon, \alpha},
$
from which follows that the WPB is also {\it complete} since the original 
set of eigenstates is a complete one.

It has been already pointed out that the division into bands can be exploited 
to optimize the basis set to the particular physical problem. A typical choice 
of the energy bands is to take $\epsilon^\alpha_n=E_F$ for a cetrain $n$ and
all $\alpha$, where $E_F$ is the Fermi energy of the system. This way the ground-state 
is described by occupying all of the WPs in the bands below $E_F$. This means 
that we need to consider only few WP or electrons even though we are describing 
the {\it local} ground state properties of the infinite many-electron system 
exactly (see \cite{EPAPS}). Similarly, the non-equilibrium state is obtained by imposing 
different effective Fermi energies for WPs with different values of $\alpha$.

We will now demonstrate the use of the WPB on several examples from 
two rapidly developing areas of condensed matter physics - time-dependent 
and/or {\it ab initio} simulations in quantum transport, and spin 
accumulation due to spin-orbit coupling in 2D systems.

{\bf Time-dependent quantum transport}.
Understanding the quantum transport of charge through nanojunctions made of individual 
atoms or molecules will be essential for progress in nanoelectronics. 
Due to the short spatial scale and short times involved it is clear that transient 
phenomena play an important role in understanding the functionality of nanodevices.
At the same time, it has been recognized that the correct treatment of interactions 
demandes a time-dependent formulation of the density- or current-density functional 
theory\cite{Koentopp08}. While several exact methods have been put 
forward\cite{Kurth05,Stefanucci04,Sai07,Burke05}, due to their inherent complexity, they give 
restricted insight into the processes involved. Here we show that the WPB can 
provide this insight in an elegant fashion, as well as quantitative results 
for transient times, oscillations or steady-state current.

\begin{figure}[t]
\includegraphics[width=9cm]{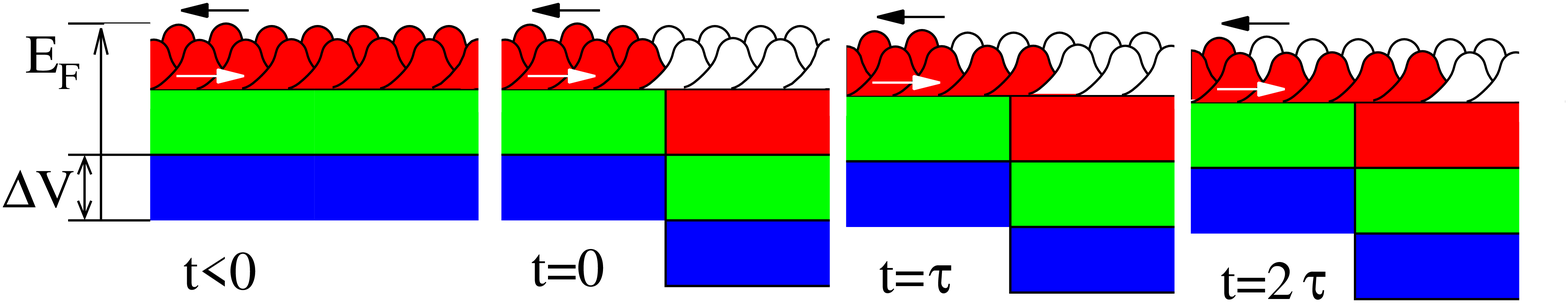}
\includegraphics[width=8cm]{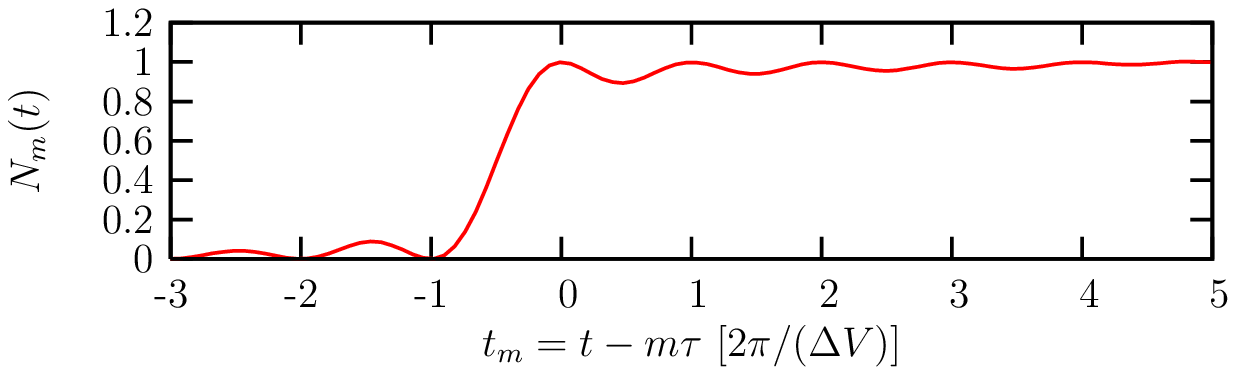}
\caption{Abrupt switching on of the bias in a 1D wire. 
In response to the bias $\Delta V$, the right-going WPs for $x>0$ (white, the previously 
unoccupied band) start to fill the WPs from the left and the occupied left-going 
WPs for $x<0$ become empty. The finite extent of each WP causes oscillations, 
with period $\tau$, of the resulting occupancy $N_m(t)$ and hecne the current 
measured at fixed $x_m=v_F m \tau$ (below).} \label{fig-2}
\end{figure}

As an example let us start with the a 1D electron gas in which at time 
$t=0$ a finite potential difference is applied (Fig.~\ref{fig-2}). 
Anticipating the application of the bias $\Delta V$, we split the occupied part 
of the spectrum of a Hamiltonian for free electrons into occupied bands
$0$ to $\Delta V$, $\Delta V$ to $E_F-\Delta V$ and $E_F-\Delta V$ to $E_F$ (we will
refer to the last band as $n=a$, i.e. the {\it active} band), and
two unoccupied bands $E_F$ to $E_F+\Delta V$ and $E_F+\Delta V$ to $\infty$.
The energy-normalized eigenstates are the
plane-waves $\braket{x}{\epsilon,\alpha} = e^{i\alpha k x}/\sqrt{2\pi k}, 
\quad k=\sqrt{2\epsilon}$, and $\alpha=\pm$ for right- and left- going 
states respectively. The resulting WPs $\braket{x}{n,m,\pm}$, obtained 
according to the Eqs.~\ref{eq-f2},\ref{eq-f3}, are examples of the bulk WPs 
mentioned above and 
are identical to the WPs employed by Martin and Landauer in thier analysis 
of quantum noise\cite{Martin92}. Due to the hopping of electrons between 
the WPs in time $\tau$, the current at the position of the $m$-th WP carried 
by electrons in the {\it active} band is in general given as $I(t)=N_m(t)/\tau$, 
where $N_m(t)$ is the occupation of the $m$-th WP.

Switching on the bias $\Delta V$ at $x=0$ and $t=0$ will energetically align 
WPs from the highest occupied, band localized in $x<0$, with the WPs from 
the lowest unoccpied band and localized in $x>0$. A transient phenomenon 
for time $t\sim 2\pi/E_F \leq \tau=2\pi/\Delta V$, which needs to be analyzed 
by performing a time-dependent simulation, will be related to dynamics 
of those occupied WPs that had for $t<0$ nonzero amplitude for both $x<0$ and $x>0$. 
After that the time-dependent many-electron dynamics for $x>0$ will result in a 
train of right-going scattering orthogonal WPs within the {\it active} band
\begin{equation}
\bra{x} a,l,+;t)=\int_{E_F - \Delta V}^{E_F} \frac{d\epsilon}{\sqrt{2\pi k \Delta V}} t(k) 
e^{i(kx-\frac{1}{2}k^2 (t+ l \tau))},
\end{equation}  
occupied for $l=0,-1,-2,...$, where $t(k)$ is the transmission amplitude for the 
applied step potential. The occupation $N_m(t)$ of the $m$-th bulk WP due to 
this train is $N_m(t) = 2 \sum_{l=0}^{-\infty} |\left(a,l,+;t \right. \ket{a,m,+} |^2$ 
which, after substituting the above expressions, gives finally 
\begin{equation} 
	I_m(t) = \frac{2}{\tau} \int \int_{E_F-\Delta V}^{E_F} d \epsilon d \epsilon'
        t^*(\epsilon') t(\epsilon) F_{t-m\tau}(\epsilon'-\epsilon), \label{eq-td-Landauer}
\end{equation}
where $F_t(\omega) =  (\Delta V)^{-2} \sum_{l=0}^{+\infty} 
\exp\{- i \omega (l\tau - t)\}$. In fact, this result is equally valid for 
abrupt swithing in 1D wire with an {\it arbitrary scattering potential} and it 
represents a generalisation of the Landauer formula for non-linear time-dependent 
response to abrupt switching on. In the long time limit we have 
$F_t(\omega) \rightarrow \delta(\omega)/(\Delta V)$ and we recover the 
non-linear Landauer formula $I= \int_{\Delta V} |t(\epsilon)|^2 d \epsilon / \pi$ .

More specifically for the 1D wire case, we can put $t(k)=1$ and perform the integration 
with the result $N_m(t) = 4 \sin^2(\Delta V t/2)/(\Delta V)^{2} \sum_{l=0}^{-\infty} 
\{(l-m)\tau + t \}^{-2}$. In the Fig.~\ref{fig-2} we show this result, calculated 
by taking the first 10 terms of this series, i.e. accounting for 10 WPs, for which 
we get a well converged answer. The relaxation to the steady-state current 
is characterised by oscillations with period $\tau$, in agreement with calculations 
based on non-equilibrium Greens functions within a wide band model\cite{Stefanucci04}. 

The WPB-based picture offers a natural framework for the memory-loss 
theorem\cite{Stefanucci04,Stefanucci07} stating the independence of the steady state 
on the transient changes in external potential. Indeed, from the moment 
when the potential attains its long-time static form, it takes only a finite time 
until the WPs experiencing the transient potential leave the scatterer into the bulk, 
never to return. After that the occupancies of all the WPs inside this region 
are determined by the scattering of the bulk WPs within the long-time static potential. 

Our treatment here also indicates that the WPB representation can be used 
to perform numerical {\it ab initio} time-dependent simulations 
within the TDDFT framework, i.e. accounting for time-dependent self-consistent field
in the scattering region. The time-evolution of the bulk WP as they enter 
the scattering region needs to be done numerically, but as soon as the scattered 
WP leaves this region, by expanding it into few bulk WPs one can perform 
its time evolution algebraically in a closed form. The density, current density 
or any other many-electron property is obtained by summing contributions from 
all stroboscopic images of the propagated WP. While the WPs will typically
extend over several atomic distances, relatively few of them will be needed 
to compute local properties close to the scattering reagion, i.e. for a 
jellium model of a sodium mono-atomic wire with one atom missing 
(creating a gap and hence depletion of charge and corresponding Friedel oscillations) 
is well converged to the exact density of an infinite system with the gap using 
about 20 occupied WPs~\cite{EPAPS}. Detailed implementation of at the self-consistent mean-field (TD DFT) methodology 
will be reported elsewhere\cite{Corsetti08}.

{\bf Edge-induced spin Hall effect}. It has been recently shown that the interplay 
between nonzero Rashba-Bytchkov spin-orbit (SO) coupling, the scattering off the edge 
and nonzero electric current along this edge leads to a universal spin polarization 
localized close to the edge of the 2D gas in GaAs quantum wells\cite{Reynoso06,Zyuzin07}. 
In parallel, several other authors~\cite{Bellucci06,Xing06,Hattori06}
considered the spin-orbit (SO) coupling due to nonzero gradient in potential in-plane, 
$
	{V}_{SO} = - \alpha_E \left[ \hat{\bf \sigma} \times
        \nabla V(\bf r) \right] \cdot \hat{\bf p},
$
where $\alpha_E$ is the strength of the SO coupling, $\hat{\bf \sigma}$ is the operator of spin, 
$V(\bm r)$ is the confining potential at the edge and $\hat{\bf p}$ the momentum operator\footnote{
We will use the effective atomic units a.u.$^*$, where $m_{eff}=\hbar=e^2/\epsilon_r=1$ with 
$m_{eff}$ and $\epsilon_r$ being the electrons effective mass and static dielectric constant 
for GaAs.}. The edge-SO scattering, analogous to the mechanism behind impurity scattering in 
the bulk of the 2D gas, seems to lead to effects similar to the Rashba-Bytchkov 
mechanism. 

Both of these effects can be understood and analyzed within the WPB 
description, but here we concentrate on the edge-SO scattering. We consider 
a 2D electron gas confined in the $xy (x>0)$ half-plane, with its edge being 
described by a model potential $V(\bm r)=W\theta(-x)$ where $\theta$ is the
step function. This model is appropriate for typical doping densities 
$n\sim 10^{12}$cm$^{-2}$ where the Fermi wavelength $\lambda_F\sim20$nm
is much larger than atomic spacing, principally determining the abruptness 
of the edge. The current is imposed in the $y$ direction. 
Fourier transforming $y\rightarrow k_y$, the SO term takes the form 
${V}_{SO}=\alpha_E \hat{\sigma}_z W \delta(x) k_y$, i.e.
electrons with up and down spins in the $z$ direction experience different 
scattering potential at the edge. For each $k_y$ we construct a WP, localized 
in the $x$ direction and constructed from the eigenstates of a bulk 2D electron gas. 
If we time-propagate an initial WPs with an average $k_x$ pointing 
towards the edge and identical for both up and down spin states (left-going WP), 
the reflected WPs for up and down spins will have two different phase shifts 
$\phi_{\uparrow/\downarrow}$, and hence a {\it mutual spatial shift} $l_S$ with respect 
to one other. For the model described here the shift, 
calculated from scattering-states' phase shift is
\begin{equation} 
	l_S = \langle \D{~}{k_x}(\phi_{\uparrow}-\phi_{\downarrow}) \rangle = \label{eq-10}
		 - 4 \alpha_E - 8 (2W - \langle e \rangle) \alpha_E^3 + \mathcal{O}( \alpha_E^4 ),
\end{equation}
where the averaging is over the energy band of the considered WP and $e=(k_x^2+k_y^2)/2$. 
We know that WPs separated by the time-step $\tau$ are orthogonal and 
we may place one electron in each WP. The non-equilibrium situation can be set in the 
standard fashion: occupying the WPs with $k_y>0$ up to $E_F+\Delta V$ and those WPs with 
$k_y<0$ only up to $E_F$. Deep inside the 2D bulk this WPs' shift will not contribute 
to any spin polarization because a series of occupied WPs within each band gives homogeneous 
density. However, since the up- and down-spin WPs are shifted, this shift must be directly 
related to the spin accumulation close to the edge so that to first order in $\alpha_E$
\begin{equation} 
	n_\uparrow-n_\downarrow \sim \int_{occ} \frac{dk_y}{2\pi} l_S n(k_y)
 	\sim -\frac{2 \alpha_E}{\pi^2} \sqrt{2E_F} \Delta V, \label{eq-12}
\end{equation}
where $n(k_y)=\sqrt{2E_F-k_y^2}/\pi$ is the number of initial WPs with momentum $k_y$.
The dependence on the magnitude of the confinement, $W$ comes only in the 
3th order, which follows from Eq.~\ref{eq-10} and \ref{eq-12} 
\begin{equation} 
	\D{~}{W}(n_\uparrow-n_\downarrow) = -\frac{8 \alpha_E^3}{\pi^2} \sqrt{2E_F} \Delta V, \label{eq-14}
\end{equation}
and hence the actual magnitude of the confinement potential is rather unimportant.
Both of the results, Eqs.~\ref{eq-12} and \ref{eq-14}, agree very well with more involved and 
exact Green's function based treatments which will be reported elsewhere\cite{BokesSpin08},
and demonstrate the usefulness of the WPB concept not only for qualitative but 
also for reliable quantitative estimates.

It is interesting to compare the edge-SO scattering with the Rashba-Bytchov mechanism.
The latter gives\cite{Zyuzin07} 
$	n_\uparrow-n_\downarrow = -\alpha_R^2 (2E_F)^{-3/2}\Delta V /(12 \pi^2) $,
where $\alpha_R$ is the strength of the Rashba coupling; in the 2D GaAs systems it attains 
values\cite{Sih05} $\alpha_R \sim 1.8 \times 10^{-10}$eV cm $= 1.55 \times 10^{-2}$a.u.$^*$.
On the other hand, the estimates for $\alpha_E$ in GaAs quantum wells give\cite{Engel05}
$\alpha_E \sim 5.3 \AA^2 =5.53 \times 10^{-4}$a.u.$^*$. The smallness of both $\alpha_E$ and $\alpha_R$
justifies the lowest order expansions used above. Finally, taking for the Fermi energy, 
$E_F=36$meV$=3.01$a.u.$^*$ corresponding to densities $n\sim 10^{12}$cm$^{-2}$
we find that the Rasba-mechanism is three orders of magnitude smaller than the edge 
spin-orbit scattering. In principle this might change at very low densities since the 
Rashba-mechanism increases while the edge SO scattering decreases with decreasing
the Fermi energy (or density) but for such low densities the behavior will 
be dominated by localization and interactions effects.

In conclusion, our stroboscopic wavepacket basis permits both physical understanding and quantitative predictions to be obtained for a variety of non-equilibrium processes in which an extended system of electrons is subject to time-evolution while being coupled to bulk reservoirs.  The stroboscopic construction permits the time-evolution of the system to be described straightforwardly, while the energy-localisation of the wavepackets within precise energy bands ensures that the Pauli principle is properly respected in coupling to the reservoirs.

The authors acknowledge fruitful discussions with Matthieu Verstraete.
This work was funded in part by the EU's Sixth Framework Programme through the Nanoquanta Network of Excellence (NMP4-CT-2004-500198).

\bibliography{WavePacket,SpinHall,ElectronicStructure,TransportTheory}
\bibliographystyle{prsty}

\end{document}